\documentclass{moriond}

\def\be{\begin{equation}}
\def\ee{\end{equation}}
\def\bea{\begin{eqnarray}}
\def\eea{\end{eqnarray}}

\newcommand{\Photo}{}

\begin{document}
\vspace*{4cm}
\title{NEW SIGNATURES OF DIRAC NEUTRALINO DARK MATTER}

\author{M. GOODSELL$^{1}$,  S. KRAML$^{2}$, 
\underline{H. REYES-GONZ\'{A}LEZ}$^{3,4}$, S. WILLIAMSON$^{5}$}

\address{
$^{1}$Laboratoire de Physique Th\'eorique et Hautes Energies (LPTHE), UMR 7589, Sorbonne Universit\'e et CNRS, 4 place Jussieu, 75252 Paris Cedex 05, France
\\
$^{2}$Laboratoire de Physique Subatomique et de Cosmologie (LPSC), Université Grenoble-Alpes, CNRS/IN2P3, 53 Avenue des Martyrs, F-38026 Grenoble, France
\\
$^{3}$Department of Physics, University of Genova, Via Dodecaneso 33, 16146 Genova, Italy
\\
$^{4}$INFN, Sezione di Genova, Via Dodenasco 33, I-16146 Genova, Italy
\\
$^{5}$Institute for Theoretical Physics, Karlsruhe Institute of Technology, 76128 Karlsruhe, Germany}

\maketitle\abstracts{%
Supersymmetric dark matter has been studied extensively in the context of the MSSM, where gauginos have Majorana masses.  Introducing Dirac gaugino masses, we obtain an enriched phenomenology from which considerable differences in, e.g., LHC signatures can be expected. Concretely, in the Minimal Dirac Gaugino Model (MDGSSM) we have an electroweakino sector extended by two extra neutralinos and one extra chargino. The bino- and wino-like states bring about small mass splittings leading to the frequent presence of scenarios with Long Lived Particles (LLPs). In this contribution, we delineate the parameter space of the electroweakino sector  of the MDGSSM, where the lightest neutralino is a viable dark matter candidate that escapes current dark matter direct detection.  We then focus on the allowed regions that contain LLPs and confront them against the corresponding LHC searches. Finally, we discuss the predominant case of long-lived neutralinos, to which no search is currently sensitive.}

\section{Introduction.}
%Most SUSY searches at the LHC are optimised for the Minimal Supersymmetric Standard Model (MSSM). However, 
Supersymmetric (SUSY) models with Dirac instead of Majorana gaugino masses can have very distinctive features. In particular, the phenomenology of neutralinos and charginos (electroweakinos or EW-inos) in Dirac gaugino models is quite different from  that of the Minimal Supersymmetric Standard Model (MSSM). The EW-ino sector of the Minimal Dirac Gaugino Supersymmetric Standard Model (MDGSSM) comprises six neutralinos and three charginos, as compared to four and two, respectively, in the MSSM.  More concretely, one obtains pairs of bino-like, wino-like and higgsino-like neutralinos, with small mass splittings {\it within} the bino (wino) pairs induced by the couplings $\lambda_S$ ($\lambda_T$) between the singlet (triplet) fermions with the Higgs and higgsino fields.  This small mass splittings can lead to long-lived charginos and/or neutralinos. 

The MDGSSM can be defined as the minimal extension of the MSSM allowing for Dirac gaugino masses. We add one adjoint chiral superfield for each gauge group: one singlet, one triplet and one octet.  We also assume that there is an underlying R-symmetry that prevents R-symmetry-violating couplings in the superpotential and supersymmetry-breaking sector, \emph{except} for an explicit breaking in the Higgs sector through a (small) $B_\mu$ term. 
Here, we concentrate on the EW-ino sector of the MDGSSM. After electroweak symmetry breaking, we obtain 6 neutralino and 3 chargino mass eigenstates. 
The neutralino mass matrix  ${\cal M}_N$ in the basis $(\tilde{B}', \tilde{B}, \tilde{W}'^0,\tilde{W}^0, \tilde{H}_d^0, \tilde{H}_u^0)$ is given by 

\begin{equation}
\begin{array}{rc@{\,}c@{\,}l}
%\begin{eqnarray}
\label{eq:NeutralinoMassMatrix} 
{\cal M}_N =  \left(\begin{array}{c c c c c c}
0  & m_{DY} & 0     & 0     &  -\frac{ \sqrt{2} \lambda_S }{g_Y} z_{ss} &   - \frac{ \sqrt{2} \lambda_S }{g_Y} z_{sc} \!\! \\
m_{DY} & 0  & 0     & 0     & -z_{sc} &   z_{ss}  \!\! \\
0     & 0     & 0  & m_{D2} & - \frac{ \sqrt{2} \lambda_T  }{g_2} z_{cs} & - \frac{ \sqrt{2} \lambda_T  }{g_2} z_{cc}  \\
0     & 0     & m_{D2} & 0   &  z_{cc} & - z_{cs}  \\
\!\! -\frac{ \sqrt{2} \lambda_S }{g_Y} z_{ss} & -z_{sc} & \!\! -\frac{ \sqrt{2} \lambda_T  }{g_2}z_{cs} &  z_{cc} & 0    & -\mu \\
\!\! -\frac{ \sqrt{2} \lambda_S }{g_Y} z_{sc} &  z_{ss} & \!\! -\frac{ \sqrt{2} \lambda_T  }{g_2} z_{cc} & -z_{cs} & -\mu & 0    \\
\end{array}\right), \!\!\!
  \end{array}
%\end{eqnarray}
\end{equation}

%{\small{
%\begin{equation}
%\begin{array}{rc@{\,}c@{\,}l}
%%\begin{eqnarray}
%\label{eq:NeutralinoMassMatrix} 
%&{\cal M}_N = \\ \nonumber
%&\!\!\! \left(\begin{array}{c c c c c c}
%0  & m_{DY} & 0     & 0     &  -\frac{ \sqrt{2} \lambda_S }{g_Y}m_Z s_W s_\beta &   - \frac{ \sqrt{2} \lambda_S }{g_Y}m_Z s_W c_\beta \!\! \\
%m_{DY} & 0  & 0     & 0     & -m_Z s_W c_\beta &   m_Z s_W s_\beta  \!\! \\
%0     & 0     & 0  & m_{D2} & - \frac{ \sqrt{2} \lambda_T  }{g_2}m_Z c_W s_\beta & - \frac{ \sqrt{2} \lambda_T  }{g_2}m_Z c_W c_\beta  \\
%0     & 0     & m_{D2} & 0   &  m_Z c_W c_\beta & - m_Z c_W s_\beta  \\
%\!\! -\frac{ \sqrt{2} \lambda_S }{g_Y}m_Z s_W s_\beta & -m_Z s_W c_\beta & \!\! -\frac{ \sqrt{2} \lambda_T  }{g_2}m_Z c_W s_\beta &  m_Z c_W c_\beta & 0    & -\mu \\
%\!\! -\frac{ \sqrt{2} \lambda_S }{g_Y}m_Z s_W c_\beta &  m_Z s_W s_\beta & \!\! -\frac{ \sqrt{2} \lambda_T  }{g_2}m_Z c_W c_\beta & -m_Z c_W s_\beta & -\mu & 0    \\
%\end{array}\right), \!\!\!
%  \end{array}
%%\end{eqnarray}
%\end{equation}}}

\noindent
%where  $s_W=\sin\theta_W$,  $s_\beta=\sin\beta$ and $c_\beta=\cos\beta$;
where $z_{ss}=m_Z\sin\theta_W\sin\beta$, $z_{sc}=m_Z\sin\theta_W\cos\beta$, $z_{cc}=m_Z\cos\theta_W\cos\beta$, $z_{cs}=m_Z\cos\theta_W\sin\beta$;
$\tan\beta=v_u/v_d$ is the ratio of the Higgs vevs; 
$m_{DY}$ and $m_{D2}$ are the `bino' and `wino' Dirac mass parameters; $\mu$ is the higgsino mass term, and 
$\lambda_S$ and $\lambda_T$ are the couplings between the singlet and triplet with the Higgs and higgsino fields respectively, corresponding to the new terms in the superpotential: $\lambda_S \mathbf{S} \, \mathbf{H_u} \cdot \mathbf{H_d} \ \mathrm{and} \ 2 \lambda_T \, \mathbf{H_d} \cdot \mathbf{T} \mathbf{H_u} $.
%\begin{equation}
%\begin{array}{rc@{\,}c@{\,}l}
%\begin{array}{>{\displaystyle}r>{\displaystyle}c>{\displaystyle}c>{\displaystyle}l}

%\begin{eqnarray}
%  W = W_{\rm MSSM} + \lambda_S \mathbf{S} \, \mathbf{H_u} \cdot \mathbf{H_d} + 2 \lambda_T \, \mathbf{H_d} \cdot \mathbf{T} \mathbf{H_u} \label{EQ:WNeq2}\,. 
%  \end{array}
%\end{eqnarray}
%\end{equation}
By diagonalising eq.~(\ref{eq:NeutralinoMassMatrix}), one obtains pairs of bino-like, wino-like and higgsino-like neutralinos,%
\footnote{For simplicity, we refer to the mostly bino/U(1) adjoint states collectively as binos, and to the mostly wino/SU(2) adjoint ones as winos.} 
with small mass splittings within the bino or wino pairs induced by $\lambda_S$ or $\lambda_T$, respectively. \
Turning to the charged EW-inos, the chargino mass matrix in the basis 
$v^+ = (\tilde{W}'^+,\tilde{W}^+,\tilde{H}^+_u)$,  $v^- = (\tilde{W}'^-,\tilde{W}^-,\tilde{H}^-_d)$ is given by:
\begin{equation}
\begin{array}{rc@{\,}c@{\,}l}
%\begin{array}{>{\displaystyle}r>{\displaystyle}c>{\displaystyle}c>{\displaystyle}l}

%\begin{eqnarray}
{\cal M}_C = 
\left(\begin{array}{c c c}
 0  & m_{D2} &\frac{ {2} \lambda_T  }{g_2} m_W c_\beta \\
m_{D2} & 0   & \sqrt{2} m_W s_\beta \\
- \frac{ {2} \lambda_T  }{g_2} m_W s_\beta & \sqrt{2} m_W c_\beta & \mu \\
\end{array}\right) \,, 
\label{eq:diracgauginos_CharginoMassarray}
  \end{array}
%\end{eqnarray}
\end{equation}
where $s_\beta=\sin\beta$ and $c_\beta=\cos\beta$. Analogously to the neutralino case, we now have wino-like $\tilde\chi^\pm$ pairs with a small splitting driven by $\lambda_T$.

In the following, we will explore the parameter space of the EW-ino sector of the MDGSSM where the lightest neutralino $\tilde\chi^0_1$ is a viable dark matter (DM) candidate. Regarding the LHC phenomenology, we will focus on the regions which include long-lived particles (LLPs). The complete study, upon which this contribution is based, can be found in \cite{Goodsell:2020lpx}.

\section{Dark matter viable scenarios.}\label{sec:DM}

We performed a Markov Chain Monte Carlo scan over the EW-ino parameters defined above to find the regions where the Lightest Supersymmetric Particle (LSP), the neutralino $\tilde\chi^0_1$, constitutes at least some percentage of the observed DM content of the Universe and escapes direct detection from the \textsc{XENON1T} experiment. The computation of the relic density and other constraints was performed with \textsc{MicrOMEGAs}~\cite{Barducci:2016pcb}; for details of the setup see \cite{Goodsell:2020lpx}.
%Points excluded by limits on $Z$ or Higgs invisible decays, 
%or by \textsc{LEP} constraints, are discarded in the scan.
%%%SK: if you mention these, you have to explain also \delta\rho and all other constraints which are used.

\begin{figure}
\begin{minipage}{0.49\linewidth}
\centerline{\includegraphics[width=0.9\linewidth]{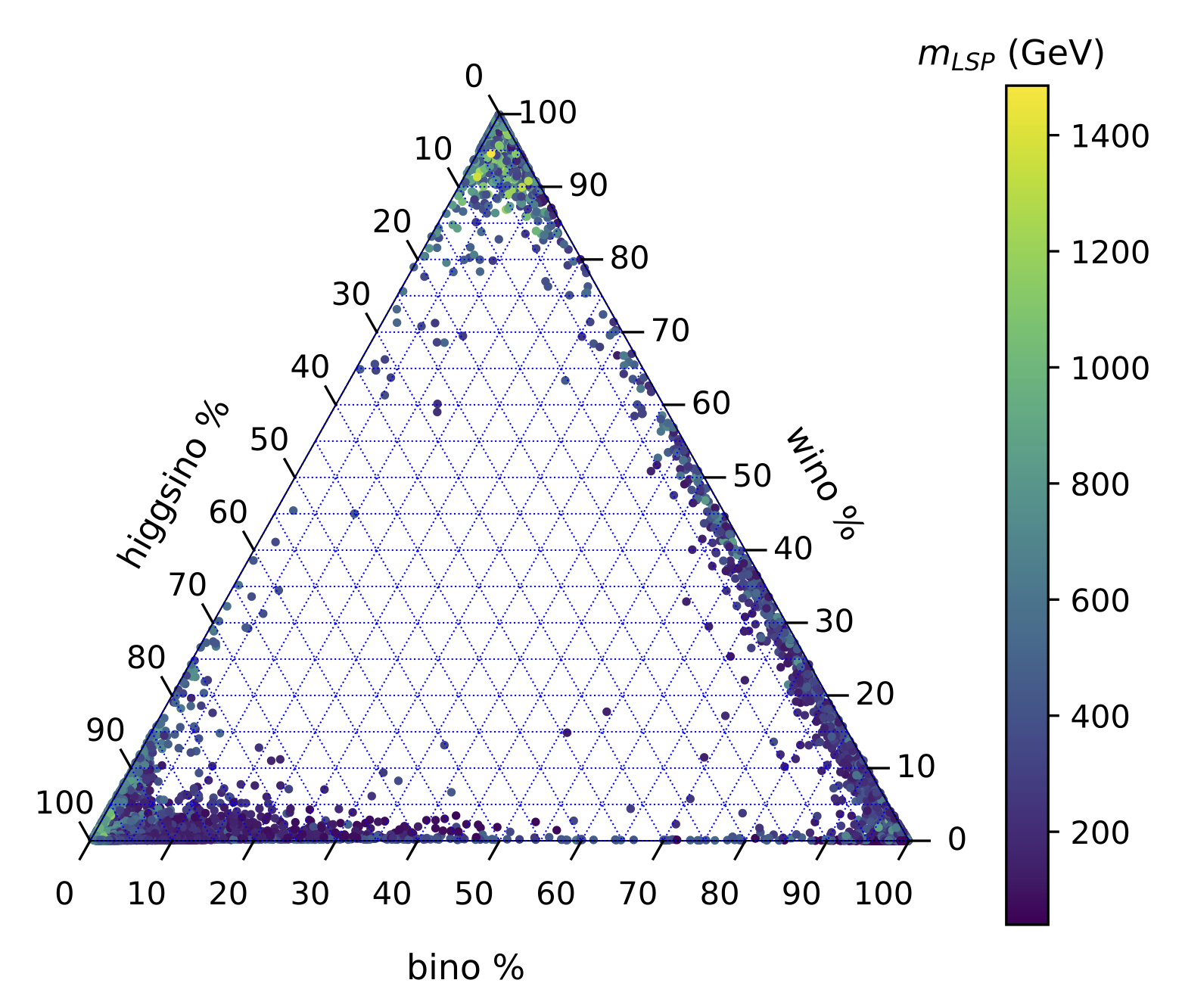}}
\end{minipage}
%\hfill
\begin{minipage}{0.49\linewidth}
\centerline{\includegraphics[width=0.9\linewidth]{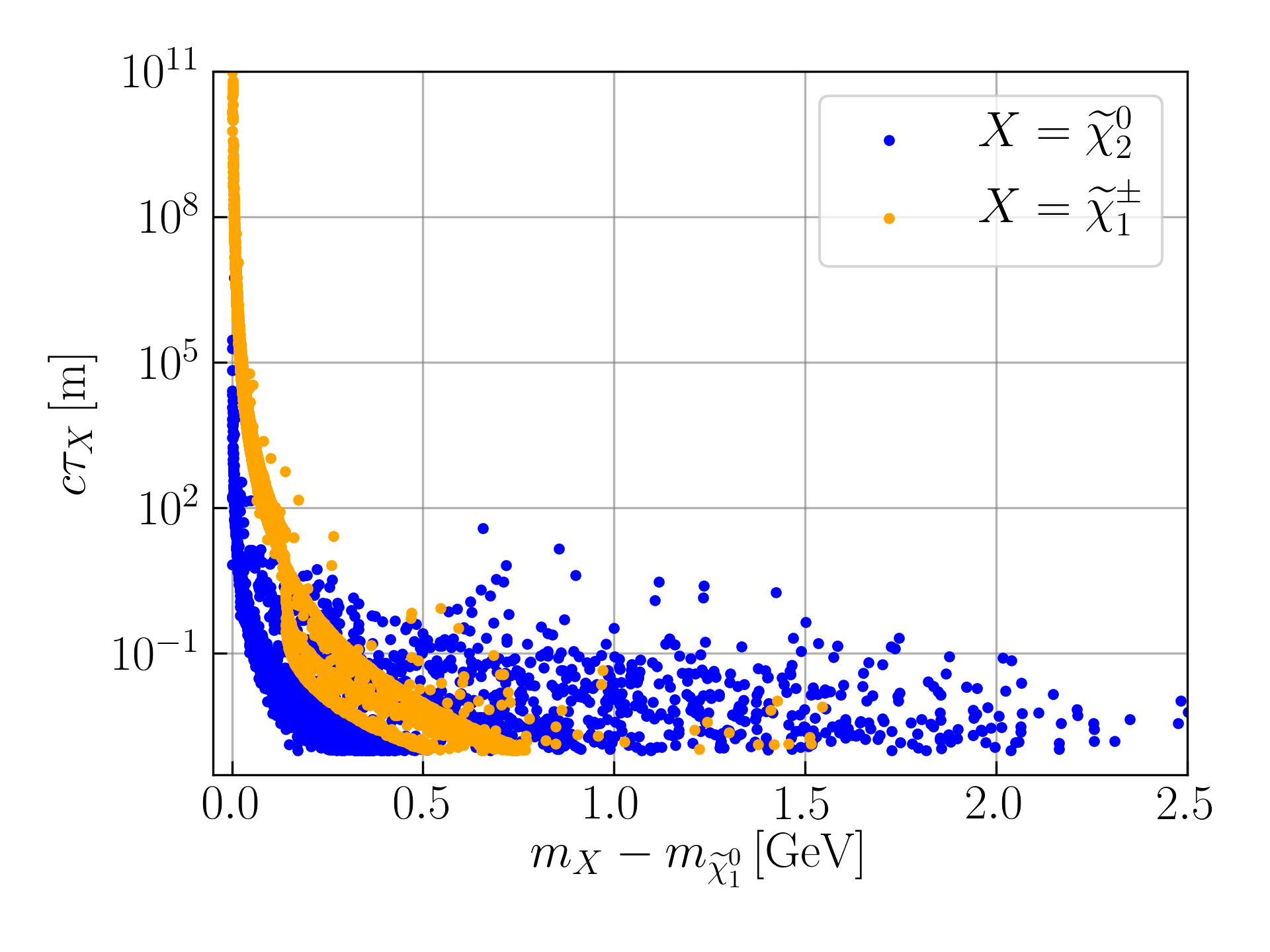}}
\end{minipage}
\caption[]{Left: bino, wino and higgsino admixtures of the LSP in the region where it makes up for  at least a part of the DM abundance. The colour denotes the mass of the LSP. Right: mean decay length $c\tau$ as a function of the mass difference with the LSP, for all points with long-lived EW-inos ($c\tau>1$~mm); blue points have a neutralino and orange points a chargino LLP.  \label{fig:DMviable}}
\end{figure}

The resulting parameter space is shown in the left panel of Figure~\ref{fig:DMviable}. The axes correspond to the bino, wino and higgsino composition of the LSP, while the color denotes its mass value. We see that cases where the $\tilde\chi^0_1$  is a mixture of all states (bino, wino and higgsino) are excluded, while cases where it is a mixture of only two states, with one component being dominant, can satisfy all constraints. Also noteworthy is that there are plenty of points in the low-mass region, $m_{\rm LSP}<400$~GeV. Interestingly, the mass differences between the LSP and the next-to-LSP (NLSP) are often so small that the NLSP (and sometimes even the next-to-next-to-LSP) becomes long-lived on collider scales, i.e.\ it has a potentially visible decay length of $c\tau>1$~mm. Concretely, about 20\% of our DM-friendly parameter points contain LLPs. The corresponding scenarios are illustrated in the right panel of Figure~\ref{fig:DMviable}, which shows the mean decay length of the LLPs as a function of their mass difference with the LSP. 

\section{Long-lived Dirac electroweakinos at the LHC.}\label{sec:LHC}

It is well known that current electroweak SUSY searches yield relatively weak bounds on general SUSY models and their constraining power on the MDGSSM is not an exception~\cite{Goodsell:2020lpx}. However, the sensitivity of LLP searches on new physics provides promising prospects for testing models with long-lived EW-inos. As seen above, such long-lived EW-inos occur naturally in the MDGSSM. 

\textit{Long-lived charginos} can be constrained by Heavy Stable Charged Particle (HSCP) and Disappearing Track (DT) searches. For the HSCP constraints we used \textsc{SModelS}~v1.2~\cite{Ambrogi:2018ujg} to evaluate the bounds from the 8~TeV and early 13~TeV (13~fb$^{-1}$) CMS analyses~\cite{Chatrchyan:2013oca,CMS-PAS-EXO-16-036}. For the DT case, we performed an interpolation of the 95\%~CL upper limits on $\sigma\times{\rm BR}$ as a function of the decay length provided by the 13~TeV ATLAS and CMS  analyses~\cite{Aaboud:2017mpt,Sirunyan:2018ldc} for 36~fb$^{-1}$ and the CMS 140~fb$^{-1}$ analysis~\cite{Sirunyan:2020pjd}. Here, $\sigma\times{\rm BR}$ stands for the direct production cross section of charginos, $\sigma(\tilde\chi^{\pm}_1\tilde\chi^{\mp}_1)+\sigma( \tilde\chi^{\pm}_1\tilde\chi^{0}_1$), times $\mathrm{BR}(\chi^{\pm}_1\rightarrow\chi^{0}_1\pi^{\pm})$. From this we computed the ratio $r$ of the predicted signal over the observed upper limit in a similar fashion as \textsc{SModelS}, considering  points with $r\ge 1$ as excluded.\footnote{DT and other lifetime dependent constraints can be treated directly in \textsc{SModelS} as of v2.0.0, released in March 2021.} 

\begin{figure}
\begin{minipage}{0.49\linewidth}
\centerline{\includegraphics[width=0.95\linewidth]{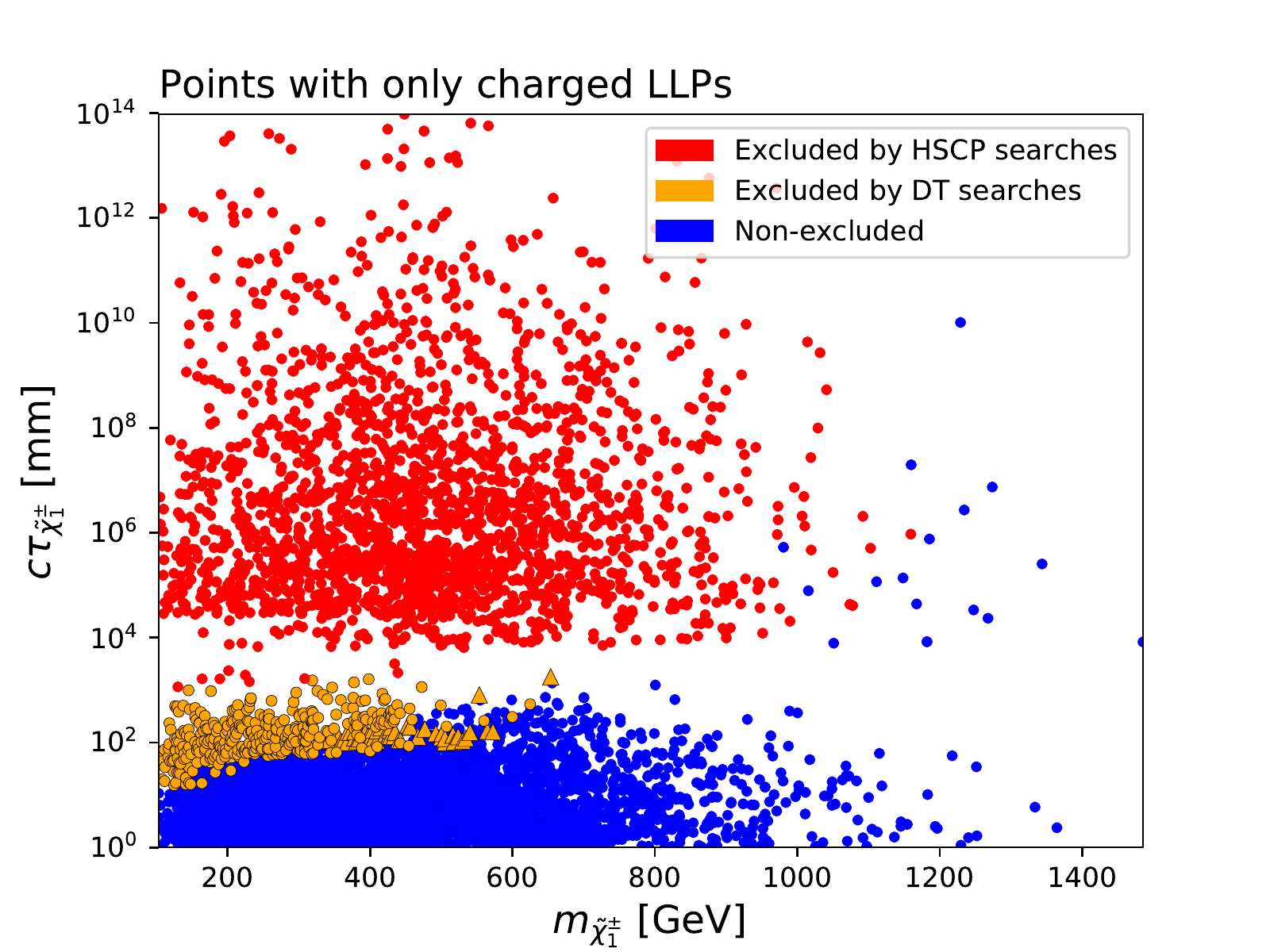}}
\end{minipage}
\begin{minipage}{0.49\linewidth}
\centerline{\vspace*{-6mm}\includegraphics[width=0.96\linewidth]{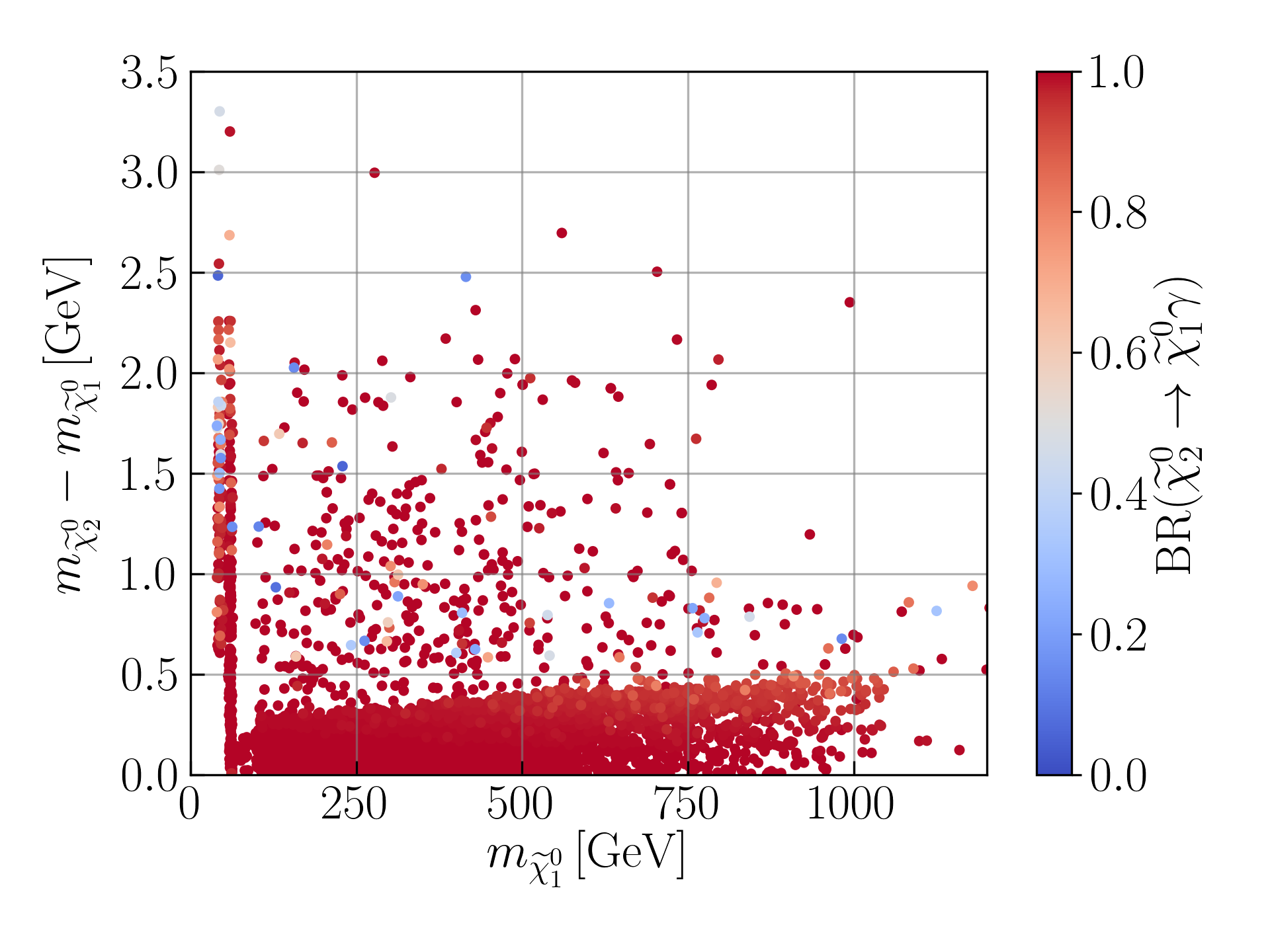}}
\end{minipage}
\caption[]{Left: Simplified-model limits for points with long-lived charginos. Red points are excluded by the HSCP searches, orange points are excluded by DT searches;  the latter are plotted as circles if excluded at $36$~fb$^{-1}$ and as triangles if excluded at $140$~fb$^{-1}$. 
Non-excluded points are shown in blue. 
Right: points with long-lived neutralinos in the plane $m_{\tilde{\chi}_1^0}$ vs.\ $m_{\tilde{\chi}_2^0}-m_{\tilde{\chi}_1^0}$; in colour the branching ratio of the loop decay $\tilde{\chi}_2^0 \rightarrow \tilde{\chi}_1^0\gamma$.}
\label{fig:LLP-LHC}
\end{figure}

The results for points with only long-lived charginos are shown in the left panel of Figure~\ref{fig:LLP-LHC} in the plane of chargino mass vs.\ mean decay length. Red points are excluded by the HSCP searches, while orange points are excluded by DT searches. The HSCP limits essentially eliminate all long-lived charginos with $c\tau_{\tilde\chi^\pm}> 1$~m up to about 1~TeV chargino mass. The exclusion by the DT searches covers $10~{\rm mm}< c\tau_{\tilde\chi^{\pm}_1}<1$~m and $m_{\tilde\chi^{\pm}_1}$ up to about 600~GeV.
The white band in-between $c\tau \approx 10^3$--$10^4$ mm corresponds to $m_{\tilde\chi^\pm_1}-m_{\tilde\chi^0_1}\approx  m_{\pi^\pm}$: 
the chargino lifetime changes abruptly when decays into pions (as well as those into muons) become kinematically forbidden, leaving decays into electrons as the only possible channel. 

%\subsection{Long-lived neutralinos.}
\textit{Long-lived neutralinos} could potentially lead to displaced vertices.  However, given the small mass differences involved, the decay products of the latter will be very soft. Furthermore, in these compressed regions, the long-lived neutralinos predominantly decays into photons via loop-induced  processes ($\tilde{\chi}^0_2 \rightarrow \tilde{\chi}^0_1 + \gamma$). The right panel in Figure~\ref{fig:LLP-LHC} shows the importance of this radiative decay in the plane of $\tilde\chi^0_1$ mass vs.\ $\tilde\chi^0_2$--$\tilde\chi^0_1$ mass difference.  As can be seen, decays into soft photons are clearly dominant. While this signature would be interesting for model 
discrimination, e.g.\ MDGSSM vs.\ MSSM, there is currently no experimental analysis which would be sensitive to it.  
% Similar would be the case at the MATHUSLA detector  \cite{Curtin:2018mvb}, which would be able to detect neutral particles that decay  $\mathcal{O}(100)$m away from the collision point at the LHC. In our case the only states that have sufficient lifetime to reach the detector have mass splittings of $\mathcal{O}(10)$ MeV (or less), and decays $\tilde{\chi}^0_2 \rightarrow \tilde{\chi}^0_1 + \gamma$ vastly dominate, with a tiny fraction of decays to electrons. However, the photons must have more than 200 MeV (or 1 GeV for electrons) to be registered at MATHUSLA. Hence, unless new search strategies are employed, our long-lived $\tilde\chi^0_2$ will escape detection. 

\section{Summary.}
We explored the parameter space of the electroweakino sector of the MDGSSM where the $\tilde\chi_1^0$ is a DM candidate in agreement with relic density and direct detection constraints. We found that a significant percentage of the viable scenarios contain long-lived neutralinos and/or charginos. For the case of long-lived charginos,  we derived strong limits by re-interpreting HSCP and DT searches from ATLAS and CMS. %, using the simplified model approach. 
However, for the case of long-lived neutralinos, we found that they %would predominantly decay to very soft photons, thus yielding a signature not covered by any experimental analysis.
predominantly yield very soft photons, which are extremely difficult to detect. 

\section*{Acknowledgments}

The work of HRG was funded by the Consejo Nacional de Ciencia y Tecnología, CONACyT, scholarship no.\ 291169. HRG is currently funded by the Italian PRIN grant 20172LNEEZ.

\end{document}